\documentclass[12pt]{iopart}
\usepackage{graphics}

\begin{document}

\title[Quantum cloning in coupled states...]{Quantum cloning in coupled
states of optical field and atomic ensemble by quasi-condensation of polaritons}

\author{A.P. Alodjants\dag, S.M. Arakelian\dag, S.N. Bagayev\ddag,
I.A.Chekhonin$\sharp$, V.S. Egorov$\sharp$ }

\address{\dag\ Vladimir State University, Department of Physics and Applied
Mathematics, ul. Gorkogo, 87, Vladimir, 600000, Russia}

\address{\ddag\ Institute of Laser Physics, Pr.Lavrentyeva 13/3,
Novosibirsk, 630090 Russia}

\address{ $\sharp$\ St.Petersburg State University, Department of Optics,
Ulyanovskyja 1, Starii Peterhof, St.Petersburg, 198504, Russia}

\eads{\mailto{alodjants@vpti.vladimir.ru}}

\begin{abstract}
We consider a new approach for storing quantum information by
macroscopic atomic excitations of two level atomic system. We
offer the original scheme of quantum cloning of optical field into
the cavity polaritons containing the phase insensitive
parametrical amplifier and atomic cell placed in the cavity. The
high temperature quasi-condensation (and/or condensation)
phenomenon for polaritons arising in the cavity under the certain
conditions is proposed for the first time.
\end{abstract}



\section{Introduction}

At present a significant progress in quantum information,
communication and especially quantum cryptography requires the new
methods and approaches in quantum information processing - see
e.g.~\cite{bib:braunstein}. Important step in this direction can
be taken with the help of novel optical devices operated with
atoms for quantum storage, memory and transmission of information.
As a rule, memory devices proposed now in atomic optics for those
purposes explore various methods of entanglement of atoms with
quantized electromagnetic field and the mapping of quantum state
of light onto the atomic ensembles~[2--6]. The principal point is
to achieve the light-atom coherence for long. For instance, in
Ref.~\cite{bib:julsgaard} atomic ensemble has been prepared as an
approximately coherent spin state with only one projection, say
$J_{x} $, having non-zero average value. Within this limit atomic
system can be described by collective magnetic momentum $J$ of the
ground state. The fidelity of up to $70\% $ and memory lifetime of
up to $4\quad msec$ have been achieved experimentally for Gaussian
light pulses in the three step scheme of passing for light-atom
interaction, and subsequent measurements in the system have been
carried out by feedback introduced into the atoms.

Another attractive and practically important opportunity to create
macroscopic coherence in atomic medium and to store coherently the
optical information as a result, is given by effect of
electromagnetically induced transparency (EIT) taking place in
three level atomic systems with so called $\Lambda $-configuration
of the energy levels. In this case strong (classical) coupling
field creates transparency window in the medium, and second weak
(quantum) probe field propagates through the resonant atomic
system with a very small absorption~\cite{bib:fleisch}. The EIT
effect is accompanied by significant reduction of observable group
velocity of propagating signal pulse, as well. Such a macroscopic
coherence as a result of the field-medium interaction has been
observed for both hot and ultracold atomic ensembles. As an
example, the authors of Refs.~\cite{bib:liu} propose to use such a
``stopped'' light for coherent (classical) storage of information
in ultracold ($T \simeq 450\quad nK$) sodium atoms near the
transition to the Bose-Einstein condensation (BEC) state. In fact,
the phenomenon is demonstrated by switching signal pulse with
delay time (i.e. storage time of light in the medium) $\tau \simeq
45\quad \mu sec$ and more. The Doppler broadening, being
limitation for the process, can be suppressed for the case. In
contrast, in Ref.~\cite{bib:phillips} the energy transfer from
light to atomic excitations for hot ($T \simeq 360\quad K$) atomic
$Rb$ vapor cell has been observed as a principal approach to
storing information.

It is important to mention that quantum memory proposals based om
EIT-type interaction in the medium leaves open principal question
about criterion for storing and transmission of quantum
information -- cf. with~\cite{bib:matsko}. Physically, quantum
properties of atomic system coupled with e.m. field, can be
described in terms of ``bright'' (optical branch) and ``dark''
(atomic branch) polaritons, being the two solutions of the problem
in the terms of linear superposition of quantized optical field
and atom excitation states -- see Refs.~\cite{bib:fleisch,bib:AVP}
for more details. But existence of polaritons is not sufficient
especially for quantum storage and transfer of information by some
quantum memory device, and specific properties of quantum state
become more important in the case -- see Ref.~\cite{bib:APA1},
cf.~\cite{bib:julsgaard}. In particular, in Ref.~\cite{bib:APA1}
we propose a new quantum non-demolition (QND) technique to store
the quantum information which is very closed to the QND
measurement procedure in quantum and atomic
optics~\cite{bib:APA2}. We have shown that atomic system under the
BEC condition satisfies formulated criteria for reasonable
efficiency of quantum information storage and transmission
simultaneously, and therefore is perspective for creating a long
lived memory. However, the atomic BEC medium is not feasible for
wide practical exploring due to extremely low temperature (about
few nanoKelvins) conditions. Therefore the study of the problem of
high-temperature quasi-condensation phenomenon for cavity excitons
and polaritons evoks a great interest~[11--13]. Although
polaritonic condensate is nonequilibrium principaly, the formation
of single (macroscopic) coherent polariton state (ground state
with momentum $\vec {k} = 0$) has been observed recently for
quantum wells placed in semiconductor
CdTe/CdMgTe-microcavity~\cite{bib:kavokin}. In
Ref.~\cite{bib:deng} the authors experimentally measure the
dispersion curve for low brunch polaritons in semiconductor
(Ga-As)-microcavity. It has been shown that the statistical
properties of excitons can be changed into the coherent ones.

Speaking more precisely such a strong coupling regime for the
cavity polaritons can be considered as a weakly interacting 2D
Bose-gas for which the Kosterlitz-Thouless phase transition occurs
at the temperatures high enough (room) due to extremely small mass
of polaritons $m_{eff} \simeq 5 \times 10^{ -
33}$g~\cite{bib:deng,bib:eastham}. The last circumstance is very
attractive and important for practical purposes and namely, for
elaboration of quantum memory devices.

Recently in Ref.~\cite{bib:averch} we have shown that a
quasi-condensation phenomenon can be also achieved for polaritons
in atomic system under the strong coupling condition for e.m.field
and two-level atoms in the cavity, described by relation

\[
\omega _{c} = \sqrt {\frac{{2\pi d^{2}\omega _{0} n}}{{\hbar} }} \gg {{1}
\mathord{\left/ {\vphantom {{1} {\tau _{coh}} }} \right.
\kern-\nulldelimiterspace} {\tau _{coh}} },
\]

\noindent where $\omega _{c} $ is so-called cooperative frequency,
$\omega _{0} $ is the frequency of atomic transition, $d$ is the
dipole momentum, $\tau _{coh} $ is the time of coherence of the
medium, $n$ is the density of atomic cloud,$\hbar $ is the Planck
constant. Within this limit the process of vacuum Rabi splitting
for the cavity modes occurs, and such a splitting corresponds to
the degeneracy of the 2D ideal polariton gas. In particular, this
takes place in a real experimental situation with the hot {\it Na}
atomic vapor cell placed into the resonator~\cite{bib:vasil}.
Moreover is also possible to achieve a true room temperatures BEC
for polaritons under the certain conditions for optical trapping
of the cavity polaritons.

In this paper we propose a new method to store quantum information
by means of quantum cloning procedure for continuous variables of
optical field in a light-atom interaction in the cavity under
condition of strong coupling regime (cf.~\cite{bib:deng}). The
high temperature quasi-condensation phenomenon for the cavity
polaritons occurs in the case (Section 2). We consider simple
model for interaction of two-level atoms with quantized e.m. field
in the cavity. $^{} $In Section 3 we examine a new scheme for
optimal quantum cloning of information in such coherent atomic
system. In Appendix we develop a Holstein-Primakoff approach to
formulate the quasi-condensation condition for macroscopic coupled
states being realized for excitations in two level bosonic system
interacting with quantum e.m. field. In conclusion we briefly
discuss an appropriate application of the scheme for the problem
of quantum memory in continuous variables.{\it }

\section{Quantum macroscopic exitations in Bose-gases and
quasi-condensation phenomenon}

To describe the interaction of two level atomic system with
quantum e.m. field we consider a localized exciton model in the
frames of the Dicke Hamiltonian for the interaction of the $N$ two
level atoms with quantized e.m. field (cf.~\cite{bib:eastham}):

\begin{equation}
\label{eq1} \fl H = \sum\limits_{\vec {k}} {E_{ph} \left( {k}
\right)\psi _{\vec {k}}^{\dag} \psi _{\vec {k}}}  +
\sum\limits_{j}^{N} {\frac{{E_{at}} }{{2}}\left( {b_{j}^{\dag}
b_{j} - a_{j}^{\dag}  a_{j}}  \right)} + \sum\limits_{\vec {k}}
{\sum\limits_{j}^{N} {\frac{{g}}{{\sqrt {N}} }\left( {\psi _{\vec
k}^{\dag} a_{j}^{\dag}  b_{j} + b_{j}^{\dag}  a_{j} \psi _{\vec
k}} \right)}}
\end{equation}

\noindent where $\psi _{\vec {k}} \left( {\psi _{\vec {k}}^{\dag}
} \right)$ is the annihilation (creation) operator for photon with
momentum $\vec {k}$; $E_{at} $ is the energy of atomic transition
between $\left| {a} \right\rangle $ and $\left| {b} \right\rangle
$ levels (we neglect here moving the atoms in the cavity), $E_{ph}
\left( {k} \right)$ defines dispersion relation for photons in the
cavity, $g$ characterizes the atom-field coupling.

For high-reflectivity mirrors of the cavity the normal (to the
plane of mirrors) component of the photon wave vector $k_{ \bot} $
is quantized, i.e. $k_{ \bot}  = \pi m \left/L_{cav}\right.$,
where $L_{cav} $ is the length of the cavity (distance between two
mirrors), $m$ is the number of modes. At the same time we have the
mode continuum in the direction being parallel to the mirrors
plane. In paraxial approximation ($k_{\parallel}  <
< k_{ \bot}  $) the dispersion relation for photon energy $E_{ph} \left( {k}
\right)$ inside of the cavity can be represented as:

\begin{equation}
\label{eq2} E_{ph} \left( {k} \right) = \hbar c\left| {\vec k}
\right| = \hbar c\sqrt {k_{ \bot} ^{2} + k_{\parallel} ^{2}}
\simeq \hbar c\left( {k_{ \bot}  + \frac{{k_{\parallel} ^{2}}
}{{2k_{ \bot} } }} \right).
\end{equation}

The polariton quasi-condensation occurs in the plane being normal to wave
vector component $k_{ \bot}  $. Physically it means that macroscopic
occupation of the state with $k_{\parallel}  = 0$ takes place in the case.
At the same time within the limit of a single cavity mode approximation (for
$k_{ \bot}  $) we can rewrite Hamiltonian (\ref{eq1}) in the form - see Appendix:

\begin{equation}
\label{eq3} H = E_{ph} \left( {k} \right)\psi ^{\dag} \psi +
E_{at} \phi _{}^{\dag}  \phi + g\left( {\psi _{}^{\dag}  \phi +
\phi _{}^{\dag} \psi _{}}  \right),
\end{equation}

\noindent where $\psi \equiv \psi \left( {\vec {k}} \right)$ is
the annihilation operator for a single mode optical field with the
$\vec {k}$ wave vector. In (\ref{eq3}) we also introduce the
collective exitation operators $\phi $, $\phi ^{\dag}$ (cf.
(A.11)):

\begin{equation}
\label{eq4} \phi = \sum\limits_{j = 1}^{N} {\frac{{a_{j}^{\dag}
b_{j}} }{{\sqrt {N}} }} , \quad \phi ^{\dag}  = \sum\limits_{j =
1}^{N} {\frac{{b_{j}^{\dag}  a_{j}} }{{\sqrt {N}} }} .
\end{equation}

The Hamiltonian (\ref{eq3}) can be diagonalized with the help of unitary
transformations

\numparts
\begin{eqnarray}
\label{eq5} \Phi _{\psi}  = \mu \psi - \nu \phi ,\\ \Phi _{\phi} =
\mu \phi + \nu \psi ,
\end{eqnarray}
\endnumparts

\noindent for annihilation operators of macroscopically populated
polaritonic modes $\Phi _{\psi}  $and $\Phi _{\phi}  $
corresponding to the upper and low brunch polaritons in the medium
respectively. The parameters $\mu $ and $\nu $ are the real
Hopfield coefficients that are determined by expressions
(cf.~\cite{bib:vasil}):

\begin{equation}
\label{eq6} \fl \quad \mu = \left( {\frac{{4g^{2}}}{{2\sqrt
{\delta _{}^{2} + 4g^{2}} \left( {\delta + \sqrt {\delta _{}^{2} +
4g^{2}}}  \right)}}} \right)^{1/2}, \quad\qquad \nu = - \left(
{\frac{{\delta + \sqrt {\delta _{}^{2} + 4g^{2}}} }{{2\sqrt
{\delta _{}^{2} + 4g^{2}}} }} \right)^{1/2}
\end{equation}

\noindent
and fulfil the condition $\mu _{}^{2} + \nu _{}^{2} = 1$; $\delta = E_{at} -
E_{ph} \left( {k} \right)$ is the energy detuning. Last parameter determines
the photonic and atomic contribution to polaritons.

Indeed, for $4g^{2} \gg \delta _{}^{2} $ from (\ref{eq6}) we have
$\mu _{}^{2} \to 1$ ($\nu _{}^{2} \to 0$), that corresponds to
exciton-like polariton $\Phi _{\phi}  $ in (5b). In another limit
when $4g^{2} \ll \delta _{}^{2} $ we have $\mu _{}^{2} \to 0$
($\nu _{}^{2} \to 1$), that corresponds to photon-like low brunch
of polaritons. From Eqs. (\ref{eq6}) shows that polariton
represents a half-matter and a half-photon ($\mu _{}^{2} = \nu
_{}^{2} = 1/2$) quasi-particle under the resonance condition for
$\delta = 0$.

The dispersion relation, i.e. the energy for the cavity polariton
of both upper brunch ($E_{\psi}  \left( {k} \right)$) and low
brunch ($E_{\phi} \left( {k} \right)$), is defined as
(cf.~\cite{bib:deng})

\begin{equation}
\label{eq7}
E_{\psi ,\phi}  \left( {k} \right) = \frac{{1}}{{2}}\left[ {E_{at} + E_{ph}
\left( {k} \right) \pm \sqrt {\left( {E_{at} - E_{ph} \left( {k} \right)}
\right)^{2} + 4g^{2}}}  \right],
\end{equation}

\noindent
where expression for $E_{ph} \left( {k} \right)$ is presented in (\ref{eq2}).

Important feature of low brunch polaritons in the cavity is
determined by the minimum of $E_{\phi}  \left( {k} \right)$ for $k
= 0$. In particular, we can fulfil both the resonance condition
$\left| {k_{ \bot} }  \right| = \left. E_{at}\right/(\hbar c)$ for
the cavity mode and the condition $k_{\parallel}  = 0$
simultaneously. The fact results in quasi-condensation of
polaritons for the case.

\section{Quantum cloning with polaritons}

Now we focus our attention on the problem of cloning optical field $\psi $
onto the atomic ensemble with the help of stationary polariton modes (5).
The principal set-up for the cloning procedure under consideration is shown
in Fig.1.

The initial radiation propagates through the phase-insensitive linear
amplifier (marked as (\ref{eq1}) in Fig.1) with the gain that is equal to two. The
linear transformation for photon annihilation operators of the signal ($\psi
$) and ancilla ($c$) modes is represented in the form:

\begin{equation}
\label{eq8} \psi = \sqrt {2} \psi _{in} + c_{in}^{\dag}  , \quad
c_{out} = \sqrt {2} c_{in}^{} + \psi _{in}^{\dag}  ,
\end{equation}

\noindent
where $c_{in} $ ($c_{out} $) is the value of operator at the input (output)
of the amplifier. For the cloning procedure we assume $c_{in} $ to be the
vacuum state.

Then, according to the scheme in Fig.1, the signal optical field is stored
in the atomic medium. The bright and dark polaritons at the output of medium
represent two clones of initial field $\psi _{in} $ as a result. One of them
reproduces the atomic excitations in the medium.

Taking into account the Eqs.(5) and Eqs.(\ref{eq6}) with $\mu _{}^{2} = \nu _{}^{2}
= 1/2$ we represent a total unitary transformation for annihilation
operators of bright (optical) $\Phi _{\psi}  $ and dark (the excitation in
matter) $\Phi _{\phi}  $ clones:

\begin{equation}
\label{eq9} \Phi _{\psi}  = \psi _{in} + \frac{{1}}{{\sqrt {2}}
}\left( {c_{in}^{\dag}  - \phi}  \right), \quad \Phi _{\phi}  =
\psi _{in} + \frac{{1}}{{\sqrt {2}} }\left( {c_{in}^{\dag}  +
\phi} \right).
\end{equation}

Let us define the quadrature components for polaritons:

\begin{equation}
\label{eq10} Q_{j}^{out} = \Phi _{j} + \Phi _{j}^{\dag}  , \quad
P_{j}^{out} = i\left( {\Phi _{j}^{\dag}  - \Phi _{j}}  \right),
\quad j = \psi ,\phi
\end{equation}

The quadrature components corresponding to Eqs.(9) evolve as:

\begin{equation}
\label{eq11}
Q_{\psi ,\phi} ^{out} = Q_{\psi} ^{in} + \frac{{1}}{{\sqrt {2}} }\left(
{Q_{c}^{in} \mp Q_{\phi} ^{in}}  \right),
\quad
P_{\psi ,\phi} ^{out} = P_{\psi} ^{in} - \frac{{1}}{{\sqrt {2}} }\left(
{P_{c}^{in} \pm P_{\phi} ^{in}}  \right),
\end{equation}
where $Q_{c}^{in} = c_{in} + c_{in}^{\dag}  $, $P_{c}^{in} =
i\left( {c_{in}^{\dag}  - c_{in}}  \right)$ are the Hermirtian
quadratures for ancilla vacuum mode at the input of the amplifier
(see Fig.1).

The expressions (9), (11) represent the desired linear
transformations for optimal cloning procedure of continuous
variables in the Heisenberg formalism~\cite{bib:cerf}. For mean
values of the quadratures (11)

\begin{equation}
\label{eq12}
\left\langle {X_{\psi ,\phi} ^{out}}  \right\rangle = \left\langle {X_{\psi
}^{in}}  \right\rangle ,
\end{equation}
where the variable $X = \left\{ {Q,P} \right\}$, $j = \psi ,\phi
$. For their variances $V_X = \left\langle {\left( {\Delta X}
\right)^{2}} \right\rangle $ following expressions are given:

\begin{equation}
\label{eq13}
V_{X}^{out} \equiv V_{X,j}^{out} = V_{X,j}^{in} + 1.
\end{equation}

The last term in Eq.(\ref{eq13}) characterizes the impossibility to perfect cloning
of quantum state.

Let us now examine optimal cloning procedure determined by
Eqs.(9), (12) for polaritons in the cavity. The quantum state of
the cavity polaritons being under condensation can be described by
the state vector (cf.~\cite{bib:eastham}):

\begin{equation}
\label{eq14} \left| {\Psi}  \right\rangle _{pol} = e^{\gamma \psi
^{\dag}}\frac{{1}}{{\sqrt {N!}} }\left( {\alpha a_{}^{\dag}  +
\beta b_{}^{\dag} } \right)^{N}\left| {vac} \right\rangle ,
\end{equation}

\noindent
where $\left| {vac} \right\rangle = \left| {0} \right\rangle _{at} \left|
{0} \right\rangle _{ph} $ denotes the total vacuum state for the cavity
photons $\left| {0} \right\rangle _{ph} $ and for the two-level atomic
system $\left| {0} \right\rangle _{at} $; $\gamma $ is the complex coherent
amplitude of the $\psi $ field. Parameters $\left| {\alpha}  \right|^{2} =
n_{a} /N$ and $\left| {\beta}  \right|^{2} = n_{b} /N$ determine a relative
population imbalance for atomic levels $\left| {a} \right\rangle $ and
$\left| {b} \right\rangle $, respectively and obey the normalization
condition $\left| {\alpha}  \right|^{2} + \left| {\beta}  \right|^{2} = 1$.

Using the polariton BEC state (\ref{eq14}) for mean values $\left\langle {X_{\psi
,\phi} ^{out}}  \right\rangle $ and variances $V_{X}^{out} $ of the output
polariton quadratures (clones) we obtain:

\begin{equation}
\label{eq15} \fl\left\langle {Q_{\psi ,\phi} ^{out}} \right\rangle
= \left\langle {Q_{\psi }^{in}}  \right\rangle \mp \sqrt {2N}
\left| {\alpha}  \right|\left| {\beta } \right|\cos\varphi , \quad
\left\langle {P_{\psi ,\phi} ^{out}} \right\rangle = \left\langle
{P_{\psi }^{in}}  \right\rangle \mp \sqrt {2N} \left| {\alpha}
\right|\left| {\beta } \right|\sin\varphi ,
\end{equation}

\begin{equation}
\label{eq16} \fl V_{Q,\psi} ^{out} = V_{Q,\phi} ^{out} =
V_{Q,\psi} ^{in} + 1 - 2\left| {\alpha}  \right|^{2}\left| {\beta}
\right|^{2}\cos^{2}\varphi , \quad V_{P,\psi} ^{out} = V_{P,\phi}
^{out} = V_{P,\psi} ^{in} + 1 - 2\left| {\alpha} \right|^{2}\left|
{\beta}  \right|^{2}\sin^{2}\varphi ,
\end{equation}

\noindent
where$\varphi = \varphi _{\alpha}  - \varphi _{\beta}  $ is the relative
atomic phase ($\alpha = \left| {\alpha}  \right|e^{i\varphi _{\alpha}
}$,$\beta = \left| {\beta}  \right|e^{i\varphi _{\beta} } $).

The expressions (14) demonstrate the deviations from ``usual''
optimal cloning procedure - cf. Eqs. (12)--(14).

To describe the quantum information processing, the fidelity
criterion is used - see e.g.~\cite{bib:braunstein,bib:cerf}. For
optimal cloning procedure of coherent optical field the fidelity
can be evaluated as:

\begin{equation}
\label{eq17}
F = \frac{{2}}{{\sqrt {\left( {1 + V_{Q}^{out}}  \right)\left( {1 +
V_{P}^{out}}  \right)}} }.
\end{equation}

Formally using the expressions (\ref{eq16}) for relevant fidelity parameter $F$
in Eq. (\ref{eq17}) one can obtain the extreme value $F = 0.8$ that is determined by
spin-squeezing for atomic system for $\left| {\alpha}  \right|^{2} = \left|
{\beta}  \right|^{2} = {{1} \mathord{\left/ {\vphantom {{1} {2}}} \right.
\kern-\nulldelimiterspace} {2}}$ and $\varphi = {{\pi}  \mathord{\left/
{\vphantom {{\pi}  {4}}} \right. \kern-\nulldelimiterspace} {4}}$.

However to perfect quantum cloning procedure we should require the
similarity for output and input quadrature components for the
scheme in Fig. 1 which is possible in low density approximation
only -- see Eq. (A.12). In particular, within the limit $\left|
{\beta}  \right| \to 0$ and $\left| {\alpha}  \right| \simeq 1$
respectively, for fidelity $F$ one can obtain a magnitude $2/3$
that corresponds to optimal quantum cloning process, i.e. optimal
transfer of quantum information, embodied in the field $\psi _{in}
$, and to coherent excitations in atomic medium -
cf.~\cite{bib:braunstein,bib:cerf}.

\section{Conclusion}

In the paper an opportunity to store quantum information in the bright and
dark polaritons arising in quasi-condensation process in the cavity is
given. We propose a special scheme for quantum cloning of signal field into
the polariton states. The relevant fidelity criterion demonstrates the
essentially nonclassical properties of storing quantum information. Such a
cloning procedure can be useful for elaborating quantum memory devices.
However, at least two additional conditions are obviously necessary to have
for those purposes. First, it is necessary to have subsequent procedure of
read-out of stored information. Second, we should have an opportunity to
retrieve or verify stored optical field.

It is evident that relevant dynamical properties of the cavity polaritons
are necessary to be included in the problem under consideration. Indeed, it
is possible to show that solution of the problem described by Hamiltonian in
Eq. (\ref{eq3}) under the (A.12) condition can also be also represented by a linear
transformation that looks like $\psi \left( {t} \right) \simeq \mu \left(
{t} \right)\psi _{} - \nu \left( {t} \right)\Phi _{\phi}  $, $\phi \left(
{t} \right) \simeq \mu \left( {t} \right)\Phi _{\phi}  + \nu \left( {t}
\right)\psi $, where $\mu \left( {t} \right),\quad \nu \left( {t} \right)$
are the time dependent transformation parameters,$\psi $ is the initial
value of optical field at the read-out stage and $\Phi _{\phi}  $
characterizes the dark polariton being one of the clones for the storing
stage. In particular, we can infer initial information from atomic
excitations, i.e. $\psi _{out} \equiv \psi \left( {t = \tau}  \right) \simeq
\Phi _{\phi}  $ at fixed time $\tau $ when $\mu = 0$ , $\nu = 1$.

In the paper we do not consider the problem of decoherence of
atomic and/or optical system that becomes important especially for
the purpose of a long lived quantum memory. Recently various
multi-pass protocols have been proposed for retrieving quantum
optical state, for verifying quantum storage state and for
increasing the fidelity, respectively~\cite{bib:julsgaard}.
Indeed, the retrieval procedure for continuous variables consists
of mapping measured light quadrature back into the atomic
quadtrature variables in the medium with certain feedback gain. In
some sense such a technique is similar to the method proposed
earlier to achieve a good QND-measurement for light quadrature
component and/or photon number -- see e.g.~\cite{bib:APA2}. The
problem of retrieval procedure for the scheme proposed by us in
Fig.1 requires detailed analysis that is not the subject of study
in the paper.

The physical scheme under discussion can be useful in problem of
quantum cryptography with continuous variables as well. In
particular, quantum cloning procedure may be used by eavesdropper
for individual attacks of communication channel --
cf.~\cite{bib:cerf}. The asymmetric cloning procedure is also
principal in the case. Necessary asymmetry can be introduced by
manipulation of detuning $\delta $ to achieve $\mu \ne \nu $ (see.
Eqs. (5)).

Finally, we would like to note that cloning procedure discussed in the paper
can be also realized in a similar way with the help of the cavity polaritons
in quantum wells on the basis of modern semiconductor technology. This
circumstance makes reasonable the method of storing quantum information from
the practical point of view.

\bigskip

\ack

This work was supported by the Russian Foundation for Basic Research (grants
No 04-02-17359 and 05-02-16576) and some Federal programs of the Russian
Ministry of Education and Science. One of the authors (A.P. Alodjants) is
grateful to the Russian private Found ``Dynasty'' and International Center
for Theoretical Physics in Moscow for financial support.

\appendix
\setcounter{section}{1}
\section*{Appendix. Coherent excitations for two-level oscillators}

Here we consider the quantum properties of macroscopic excitations
for a two level atomic system interacting with e.m. field. We use
the general Holstein-Primakoff approach for that --
cf.~\cite{bib:holstein}.

In Schwinger representation the pseudospin operators for a two level system
can be written in the form
\begin{eqnarray}
S_{x} = \frac{{1}}{{2}}\left( {a^{\dag} b + b^{\dag} a} \right),
\quad S_{y} = \frac{{i}}{{2}}\left( {a^{\dag} b - b^{\dag} a}
\right),\nonumber\\ S_{z} = \frac{{1}}{{2}}\left( {b^{\dag} b -
a^{\dag} a} \right), \quad S_{0} = \frac{{1}}{{2}}\left( {a^{\dag}
a + b^{\dag} b} \right),
\end{eqnarray}
\noindent where $a\left( {a^{\dag} } \right)$ and $b\left(
{b^{\dag} } \right)$ are the annihilation (creation) operators for
the oscillators (atoms) at the lower $\left| {a} \right\rangle $
and upper $\left| {b} \right\rangle $ levels, respectively. These
operators obey the usual Bose commutation relations; $S_{0} $ is
the operator of the total particle number. Physically expressions
(A.1) mean that in the case of BEC we can consider two {\it
macroscopically} occupied quantum modes $a$ and $b$ for internal
states $\left| {a} \right\rangle $ and $\left| {b} \right\rangle $
of atomic system instead of collective operators (\ref{eq4}) --
cf.~\cite{bib:julsgaard,bib:APA1}. In other words the condensation
(and/or quasi-condensation) phenomenon leads to the interaction of
a single quantum cavity mode and atomic ensemble coherently
similar to a single two level atom described by operators $a\left(
{a^{\dag} } \right)$ and $b\left( {b^{\dag} } \right)$.

The operators $S_{j} $ satisfy the commutation relation of the
SU(\ref{eq2})-algebra:
\begin{eqnarray}
\left[ {S_{x} ;S_{y}}  \right] = iS_{z} ; \quad \left[ {S_{z}
;S_{x}}  \right] = iS_{y} ; \quad \left[ {S_{y} ;S_{z}}  \right] =
iS_{x};\nonumber\\ \left[ {S_{j} ;S_{0}}  \right] = 0,\quad j =x,
y, z.
\end{eqnarray}

It is useful to introduce the ladder operators:
\begin{equation}
S_{ +}  = S_{x} + iS_{y} = b^{\dag} a, \quad S_{ -}  = S_{x} -
iS_{y} = a^{\dag} b,
\end{equation}
which obey the commutation relations:
\begin{equation}
\left[ {S_{ +}  ;S_{ -} }  \right] = 2S_{z} ; \quad \left[ {S_{z}
;S_{ \pm} }  \right] = \pm S_{ \pm}.
\end{equation}

For the basis of two-mode Fock state, i.e. for
\begin{equation}
\left| {s_{z}}  \right\rangle = \left| {n_{a} ,n_{b}}
\right\rangle,
\end{equation}
the $S_{z} $ and $S_{0} $ operators are diagonal with the mean
values
\begin{equation}
s_{z} \equiv \left\langle {S_{z}}
\right\rangle = \frac{{1}}{{2}}\left( {n_{b} - n_{a}}  \right),
\quad s \equiv s = \left\langle {S_{0}}  \right\rangle =
\frac{{1}}{{2}}\left( {n_{b} + n_{a}}  \right),
\end{equation}
where $n_{a} = \left\langle {a^{\dag} a} \right\rangle $, $n_{b} =
\left\langle {b^{\dag} b} \right\rangle $ are the average number
of particles at the $\left| {a} \right\rangle $ and $\left| {b}
\right\rangle $ levels. Using the expressions (A.6) it is easy to
show that
\begin{eqnarray}
S_{ +}  \left| {s_{z}}  \right\rangle = \sqrt {\left( {s + s_{z} +
1} \right)\left( {s - s_{z}}  \right)} \left| {s_{z} + 1}
\right\rangle,\nonumber\\ S_{ -}  \left| {s_{z}}  \right\rangle =
\sqrt {\left( {s - s_{z} + 1} \right)\left( {s_{z} + s} \right)}
\left| {s_{z} - 1} \right\rangle.
\end{eqnarray}

Let us introduce the spin excitation operators $\phi \left( {\phi
^{\dag} } \right)$ according to the Holstein-Primakoff
transformation as:
\begin{eqnarray}
S_{ -}  = \sqrt {2s - \phi ^{\dag} \phi}  \phi, \quad S_{ +}  =
\phi ^{\dag} \sqrt {2s - \phi ^{\dag} \phi},\nonumber\\ S_{z} =
\phi ^{ +} \phi - s.
\end{eqnarray}

The commutation relations (A.4) are satisfied when the operators
$\phi \left( {\phi ^{\dag} } \right)$ obey the following relations
for Bose-system :
\begin{equation}
\left[ {\phi ;\phi ^{\dag} } \right] = 1.
\end{equation}

In the paper we consider the limit of small number of excitations
$\left\langle {\phi ^{\dag} \phi}  \right\rangle $, i.e.:
\begin{equation}
\left\langle {\phi ^{\dag} \phi}  \right\rangle < < s,
\end{equation}
that corresponds to low density excitons approach -
cf.\cite{bib:deng}. In this case the operators $\phi ^{\dag}
\left( {\phi} \right)$ can be evaluated as (cf. Eqs.(\ref{eq4}) ):
\begin{equation}
\phi = \frac{{S_{ -} } }{{\sqrt {2s}} } = \frac{{a^{\dag}
b}}{{\sqrt {N}} }, \quad \phi ^{\dag}  = \frac{{S_{ +} } }{{\sqrt
{2s}} } = \frac{{b^{ +} a}}{{\sqrt {N}} }.
\end{equation}

Using the expressions (A.6), (A.11) it easy to present the relation (A.10)
in the form:
\begin{equation}
n_{a} > > n_{b}.
\end{equation}

Thus, within this limit the boson-like excitations (polaritons) in the
medium exhibit the quasi-condensation properties.

\newpage

\begin{figure}
\resizebox{0.7\textwidth}{!}{%
  \includegraphics{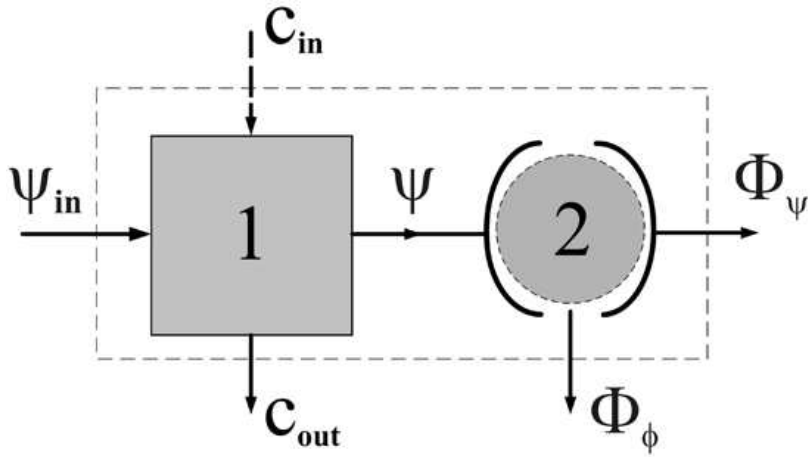}
} \caption{Scheme of quantum cloning of light onto the polaritons.
Here $\psi _{in} $ ($c_{in}^{} $) is the annihilation operator for
signal (ancilla) optical field at the input of a cloning device; we
denote 1 as a linear amplifier, 2 is the atomic cell placed in the
cavity, $\Phi _{\psi}  $ and $\Phi _{\phi}  $ are the polariton
modes after the storage procedure has occured.}
\end{figure}

\end{document}